\documentstyle[preprint,aps,eqsecnum]{revtex}
\input epsf

\newcommand{\eeq}{\end{equation}}

\newcommand{\beq}{\begin{equation}}
\newcommand{\beql}{\begin{eqnarray}}
\newcommand{\eeql}{\end{eqnarray}}

\begin{document}

\draft

\title{Spectral Statistics of Instantaneous Normal Modes in Liquids and Random Matrices}

\author{Srikanth Sastry$^1$, Nivedita Deo$^{1,3}$ and Silvio Franz$^2$}

\address{$^1$ Jawaharlal Nehru Center for Advanced Scientific Research, Bangalore 560064, India}
\address{$^2$ The Adbus Salam International Centre for Theoretical Physics,
Treieste, Italy}
\address{$^3$ Santa Fe Institute,
1399 Hyde Park Road, Santa Fe, NM 87501, USA}

\maketitle

\begin{abstract}
We study the statistical properties of eigenvalues of the Hessian
matrix ${\cal H}$ (matrix of second derivatives of the potential
energy) for a classical atomic liquid, and compare these properties
with predictions for random matrix models (RMM). The eigenvalue
spectra (the Instantaneous Normal Mode or INM spectra) are evaluated
numerically for configurations generated by molecular dynamics
simulations. We find that distribution of spacings between nearest
neighbor eigenvalues, $s$, obeys quite well the Wigner prediction $~
s~exp(-s^2)$, with the agreement being better for higher densities at
fixed temperature. The deviations display a correlation with the
number of localized eigenstates (normal modes) in the liquid; there
are fewer localized states at higher densities which we quantify by
calculating the participation ratios of the normal modes. We confirm
this observation by calculating the spacing distribution for parts of
the INM spectra with high participation ratios, obtaining greater
conformity with the Wigner form. We also calculate the spectral
rigidity and find a substantial dependence on the density of the
liquid.
\end{abstract}


\section{Introduction:}
The local topography of the potential energy surface, as characterized
by the ensemble averaged spectrum (INM spectrum) of eigenvalues of the
second-derivative matrix (Hessian) of the potential energy function,
have been studied in recent years as an approach to the analysis of
dynamics in liquids\cite{keyes,Sciortino}. Broadly, the study
of the INM spectra has been directed at analysing short time dynamics 
as in studying solvation \cite{strattcho}, and at elucidating information
about pathways to long time relaxation in the form of potential energy 
barriers {\it etc}\cite{keyes,Sciortino}. 

Considerable effort has also been dedicated to developing analytical
theories for calculating the INM spectra within an equilibrium
description\cite{xustratt,wu,stratt,mona,cava}. The approach in much
of these attempts has been to formulate the problem of calculating the
INM spectrum as an exercise in random matrix theory. If one treats the
individual elements of the Hessian matrix as independent, and
distributed according to the appropriate Boltzmann weight, then the
Hessian may be viewed as a real, symmetric random matrix with a known
distribution of matrix elements. Two properties, however, distinguish
the Hessian from the standard corresponding case treated in random
matrix theory: (i) The diagonal entries of the Hessian are related to
the off-diagonal entries by the property ${\cal H}^{\alpha \beta}_{ii}
= - \sum_{j \neq i} {\cal H}^{\alpha \beta}_{ij}$, where $i,j$ label
the particles and $\alpha, \beta$ the spatial coordinates $x, y,
z$. (ii) For liquids with short-ranged interaction potentials, the
Hessian matrix is sparse, with the fraction of non-zero entries, $p$,
depending on the system size as $p \sim 1/N$.

In view of the above considerations, it is of interest to inquire to
what extent the INM spectrum displays universal features idenitied in
random matrix theory. In this paper, we address this question by
obtaining INM spectra numerically for a model atomic liqiud that has
been studied in the context of slow dynamics in supercooled
liquids\cite{sastryNature}. The statistics we consider are the
spacing statistics between nearest neighbor eigenvalues and the 
spectral rigidity, which we explain below. 

\noindent
\section{Eigenvalues of the Hessian}
The model liquid we study is a binary mixture\cite{kob} composed of
$80 \%$ of particles of type $A$ and $20 \%$ of type $B$, interacting
via the Lennard-Jones potential, with Lennard-Jones parameters
$\epsilon_{AB}/\epsilon_{AA} = 1.5$, $\epsilon_{BB}/\epsilon_{AA} =
0.5$, $\sigma_{AB}/\sigma_{AA} = 0.8$, and $\sigma_{BB}/\sigma_{AA} =
0.88$, and a ratio of masses $m_B/m_A = 1$. Lennard-Jones reduced
units are used to report all the quantities, in terms of the $A$
particle paramters $\epsilon_{AA}$, $\sigma_{AA}$ and $m_{A}$:
temperatures as $T^{*} = k_B T/\epsilon_{AA}$, densities as $\rho^{*}
= \rho/\sigma_{AA}^3$ and Hessian eigenvalues $\lambda^{*} = \lambda
{m_A \sigma_{AA}^2 \over \epsilon_{AA}}$. Further details may be found
in \cite{sastryNature}.  Molecular dynamics simulations of the liquid
are performed at ten reduced densities $\rho^{*}$ from $\rho^{*} =
0.65 $ to $1.40$, at reduced temperature $T^{*} = 1.0$. Two hundred
sample configurations in each case are chosen from the equilibrated
trajectory for the INM analysis. For each of these configurations, the
Hessian is calculated and diagonalized numerically to obtain the
eigenvalues $\lambda_i$ as well as the eigenvectors ${\bf e}_i$. The
eigenvectors are used to calculate the localization properties of the
normal modes, {\it via} the participation ratio

\begin{equation}
P_i = \left[N \sum_{\alpha = 1}^{3N} ({\bf e}_i^\alpha.{\bf e}_i^\alpha)^2 \right]^{-1}. 
\end{equation}

The participation ratio thus defined is small (order of $1/N$) for localized modes and large (order of $1$) for extended modes. 

By constructing the histogram of eigenvalues $\lambda$ of the Hessian
for all configurations considered, we obtain the INM Density of States
(DOS) or spectrum\cite{fn1}. Figure 1 shows the INM spectrum
$D(\lambda)$ verses $\lambda$ for different densities. Note that the
DOS is very different from the Wigner semi-circle distribution,
obtained in RMM. We know from the literature\cite{bz} that the
correlation and spacing functions are universal in certain regions no
matter what the DOS is. Thus we use the corresponding statistics of
spacings between eigenvalues, Wigner-Dyson statistics, as the standard
of reference.

\noindent
\section{Unfolding The Spectrum}
The statistical analysis of the numerical data proceeds by first using
an {\it unfolding procedure}. The numerical calculation yields the
eigenvalues of the Hessian which is ordered and forms the sample
spectrum $\{\lambda_1,\lambda_2,...,\lambda_n\}$. In order to analyse
the spacing statistics, one must transform the eigenvalues $\lambda_i$
in such a way that the transformed eigenvalues $\zeta_i$ are uniformly
distributed.  That is, the spectral density $D(\zeta) = 1$. This
procedure is referred to as ``unfolding'' the
spectrum\cite{mehta,guhr}. In general such a transformation for a
spectral density function $D(\lambda)$ is most easily accomplished
through its cumulative distribution
\begin{equation}
C(\lambda) \equiv \int_{-\infty}^{\lambda} D(\lambda) d\lambda,
\end{equation}
by defining $\zeta(\lambda) = C(\lambda)$. For a discrete spectrum
such as the ones we consider here, the corresponding procedure would
be to extract from the sample spectrum the ``smooth'' part of the
``staircase'' cumulative distribution\cite{guhr}. In this work, the
procedure we adopt to estimate the smooth part of the cumulative
distribution is to evaluate the cumulative distribution for the union
of all eigenvalues $\lambda_i$ obtained for $200$ independent
configurations at each density and temperature value. Such a smooth
cumulative spectrum along with the ``staircase'' cumulative spectrum
for a single configuration, is shown in Fig. 2, and the unfolding
procedure is indicated.

\noindent
\section{Spacing Distribution}
The spacing distribution $P(s)$ for the random matrix models is
defined as the probability of finding the next nearest neighbor
eigenvalue of the spectrum to be at a distance s i.e.  $
s_i={{\lambda_{i+1}-\lambda_i}\over {\Delta}} $ where $ \Delta $ is
the mean level spacing. In the present case where we use unfolded
eigenvalues, $s_i=\zeta_{i+1}-\zeta _{i}$.  Then $P(s)= A s^{\beta}
e^{-B s^2}$ where $\beta=1$ for the orthogonal random matrix model,
which is the ``Wigner surmise''\cite{mehta}.  This spacing
distribution arises in various quantum systems which show an
underlying classically chaotic behavior e.g. quantum billiards,
quantum dots, nuclear spectra, disordered mesoscopic systems. The
system we analyze, on the other hand, is a classical liquid, with a
disordered microscopic structure. The spacing distributions obtained
are shown in Fig. 3 for three densities of the liquid. We find that to
a first approximation, the cases considered display the universal
behavior according to the Wigner surmise, with the agreement being
better for the liquid at higher densities. We note here that in
Ref. \cite{cava} a sparse random matrix is used to calculate the INM
DOS analytically for a one-dimensional system which is very different
from the usual Wigner semi-circle, and displays qualitative features
very similar to the DOS we calculate numerically here.  It would be
very interesting to see whether the correlators of the sparse random
matrix proposed in ref. \cite{cava} captures the above behaviour for
different densities for the liquid.

\noindent
\section{Spectral Rigidity}
We next study the spectral rigidity, which measures the fluctuations
of the number of eigenvalues in a window of given size as a function
of the size of the window (or equivalently, the average number of
eigenvalues expected in the window). The number fluctuations are
plotted as a function of the mean number of eigenvalues in Fig. 4 for
the same three densities as in the preceding figures. For the case of
the Poisson spectrum and the harmonic oscillator\cite{mehta} the
number fluctuations are $ <\delta N^2> \sim <N> $ and $ <\delta N^2> =
constant $, while for the Gaussian random matrix ensembles $ <\delta
N^2> \sim ln <N> $ for $ N \gg 1 $.

The number fluctuations are of the form $ <\delta N^2> \sim
N^{\gamma} $ where $\gamma(\rho)$ depends on the liquid's density.
This situation is reminescent of that found for the mobility edge in
the quantum hall effect and in the Anderson metal-insulator
transition.  A detailed analysis of this awaits future work.

\noindent
\section{Participation Ratio}
In order to get some insight into the reason for the systematically
better agreement with increasing density of the spacing distribution,
we consider the localization properties of the normal modes in the
liquid.  In the standard random matrix case for orthogonal matrices,
the eigenvectors are all extended, while from previous numerical and
analytical studies we know that a fraction of the INM eigenmodes are
localized. As described earlier, we use the participation ratio to
quantify the localization of modes, averaging over modes corresponding
to eigenvalues in each histogram bin, for the unfolded eigenvalues.
In Fig. 5, the participation ratio is plotted as a function of the
unfolded eigenvalues for $T^{*} =1.00$, for values of the density
$\rho^{*} =0.65, 1.00$ and $1.35$. One notes that for the highest density
$\rho=1.35$ the participation ratios are highest overall, while for the 
lower densities the participation ratios are quite small for a substantial 
fraction of the eigenmodes, indicating a large number of localized modes.

Next, we calculate the spacing distribution for $\rho^{*} = 1.35$ for
unfolded eigenvalues between $0.1$ and $0.6$ for which the
participation ratio is high ($> 0.75$) and relatively unchanging
(Fig. \ref{fig:prat}). The resulting spacing distribution is shown in
Fig. 6, along with the spacing distribution for the entire eigenvalue 
spectrum and the expectation based on the RMM result. The data shown 
clearly demonstrate that the spacing distribution is practically 
identical to the standard RMM result, confirming the speculation that 
the increasing fraction of localized states at lower densities are 
responsible for the deviations at these densities from the standard 
RMM result. Indeed, this observation has been used, in a different context
to locate the mobility edge in disordered systems\cite{carpena}

\noindent
\section{Conclusions}
We have presented the spacing statistics and spectral rigidity for
numerically calculated INM spectra. The spacing statistics is seen to
conform better with increasing density (at fixed temerature), with the
predictions for random matrix models. We demonstrate that the source
of deviations from RMM predictions is related to the presence of
localized instantaneous normal modes in the liquid, whose number is
greater for lower density. As the two features distinguishing the INM
spectra from the standard random matrix case are the nature of the
diagonal elements and the sparseness of the Hessian matrix, further 
understanding of the non-universality of the INM spectra are to be
sought in the manner in which these aspects affect the INM spectral 
statistics.

\noindent

\begin{flushleft}

\underline{Acknowledgements}:\\ 

We would like to thank S. Jain, V. E. Kravtsov, A. Cavagna,
I. Giardina, P. J. Garrahan and P. Carpena for very useful discussions
during the course of this work. SF thanks the Jawaharlal Nehru Center
for Advanced Scientific Research for hospitality.
\end{flushleft}

\begin{figure}
\caption{INM Density of States shown for three densities. Inset shows
the DOS on a logarithmic scale. Eigenvalues $\lambda$ are expressed in
units of ${\epsilon_{AA} \over m_A \sigma_{AA}^2}$, and $D(\lambda)$ in
units of ${m_A \sigma_{AA}^2 \over \epsilon_{AA}}$.}
\label{fig:inmdos}
\end{figure}

\begin{figure}
\caption{A portion of the staircase cumulative spectrum from a single
configuration is shown, along with the estimate of the smooth part of
the cumulative distribution, obtained from combining eigenvalues from
200 configurations, for density $\rho^{*} = 1.0$ and temperature
$T^{*} = 1.0$. The arrows indicate the mapping of any given eigenvalue
$\lambda_i$ to the unfolded eigenvalue $\zeta_i$. The inset shows the
cumulative spectra for the full range. Eigenvalues $\lambda$ are
expressed in units of ${\epsilon_{AA} \over m_A \sigma_{AA}^2}$, and 
cumulative probabilities $\zeta$ are dimensionless.}
\label{fig:unfold}
\end{figure}

\begin{figure}
\caption{The level spacing distribution $P(s)$ for densities $\rho =
0.65, 1.0, 1.35$. Level spacings $s$ are expressed in units of
${\epsilon_{AA} \over m_A \sigma_{AA}^2}$, and $P(s)$ in units of
${m_A \sigma_{AA}^2 \over \epsilon_{AA}}$.}
\label{fig:pofs}
\end{figure}

\begin{figure}
\caption{Spectral rigidity: 
Lines are fits to the form $<N^2> - <N>^2 \sim <N>^\gamma$, and the
values of $\gamma$ are $1.036, 0.85, 0.765$ respectively for $\rho^{*} = 0.65, 1.00, 1.35$.  }
\label{fig:rigid}
\end{figure}

\begin{figure}[t]
\caption{Participation ratio as a function of unfolded eigenvalues
$\zeta$ (see caption of Fig. 2).}
\label{fig:prat}
\end{figure}

\begin{figure}[t]
\caption{The level spacing distribution shown for $\rho^{*} = 1.35$ for
the full eigenvalue spectrum (open rhombs), for the range of
eigenvalues with high participation ratio ($\> 0.75$) (filled rhombs),
along with the RMM prediction. Level spacings $s$ are expressed in units of
${\epsilon_{AA} \over m_A \sigma_{AA}^2}$, and $P(s)$ in units of
${m_A \sigma_{AA}^2 \over \epsilon_{AA}}$.}
\label{fig:pofs2}
\end{figure}


\begin{thebibliography}{99}
\bibitem{fn1} Note that we have not followed the standard practing in
the INM literature of plotting the DOS against $\omega \equiv sign(\lambda)\times \sqrt{|\lambda|}$.
\end{thebibliography}
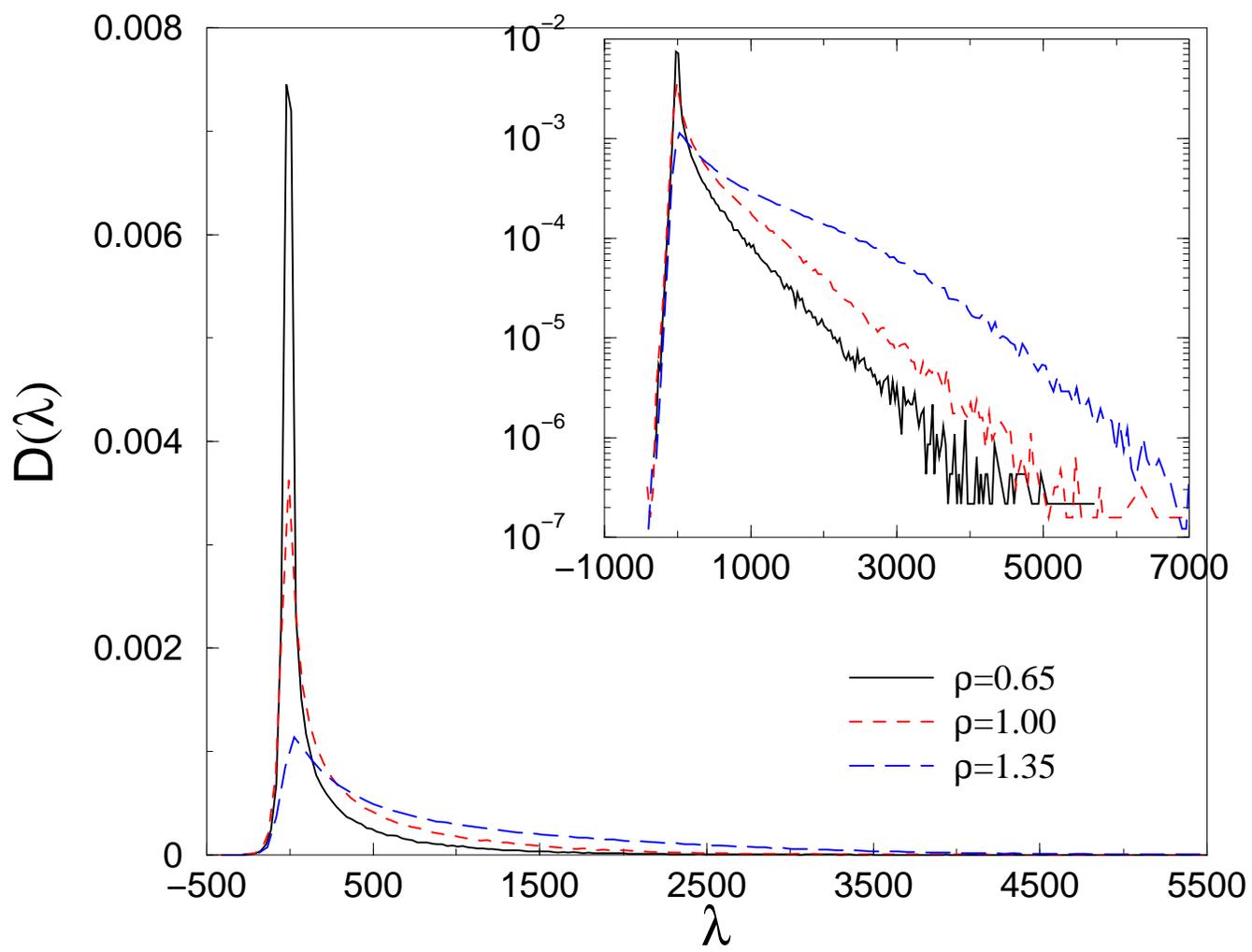
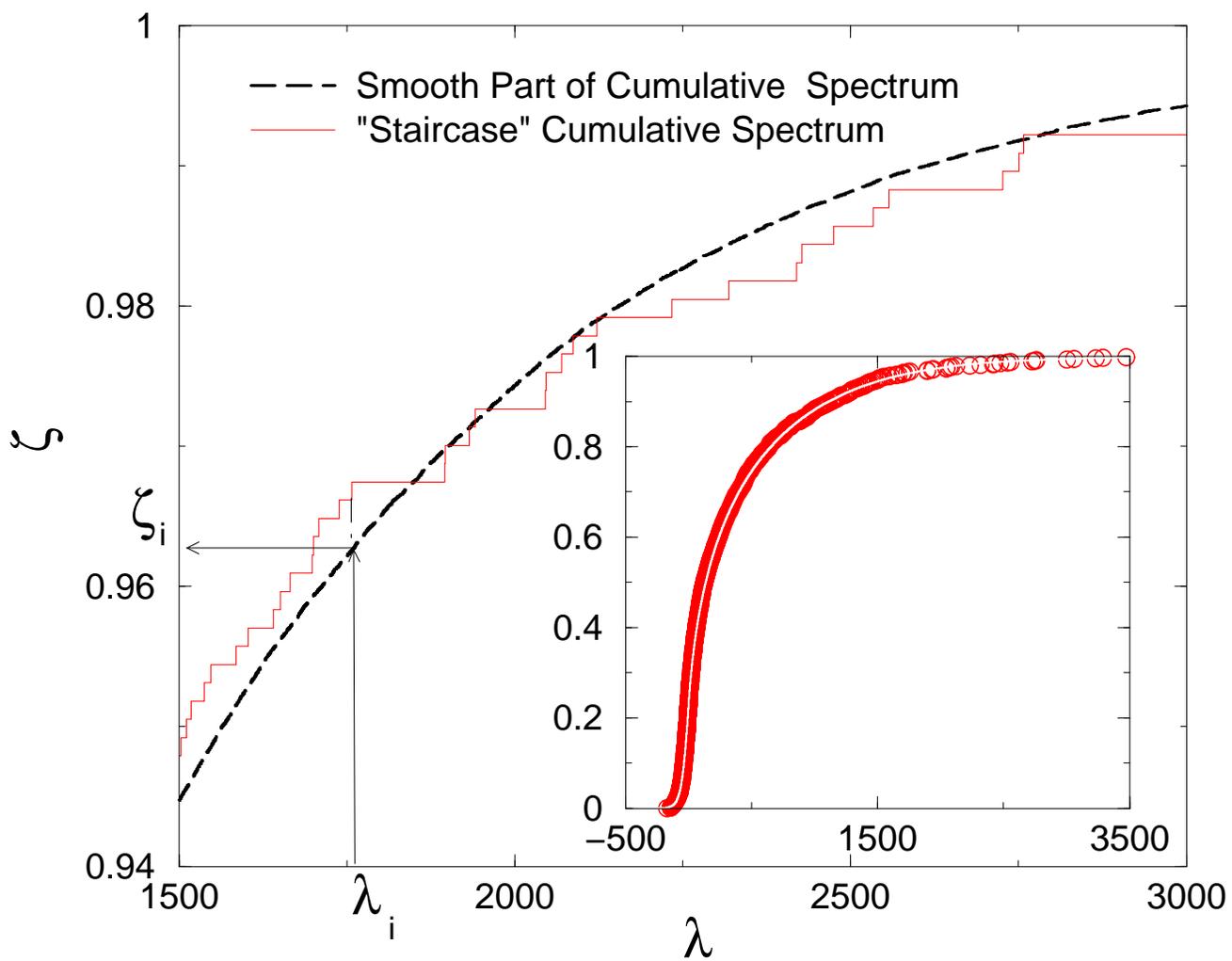
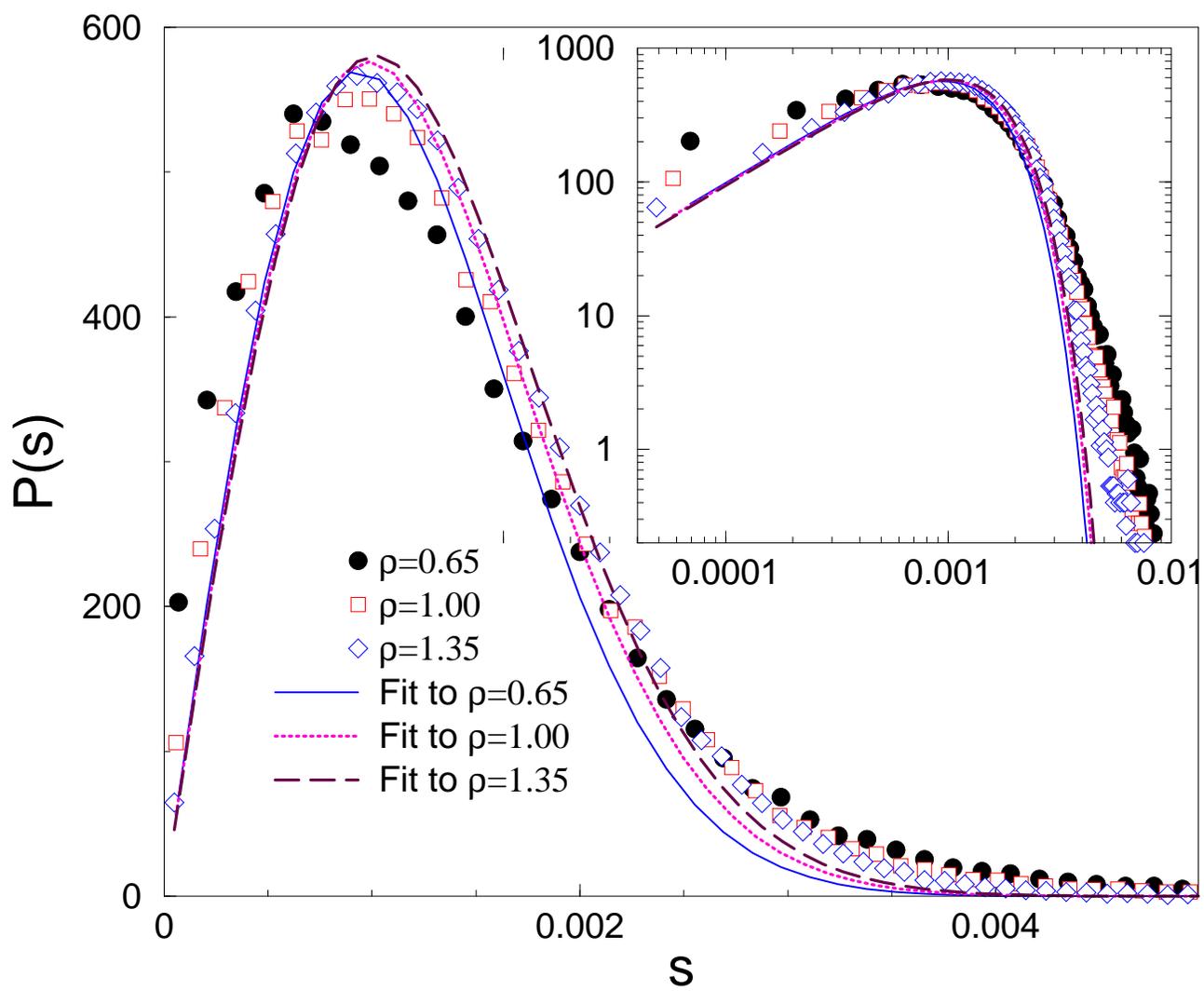
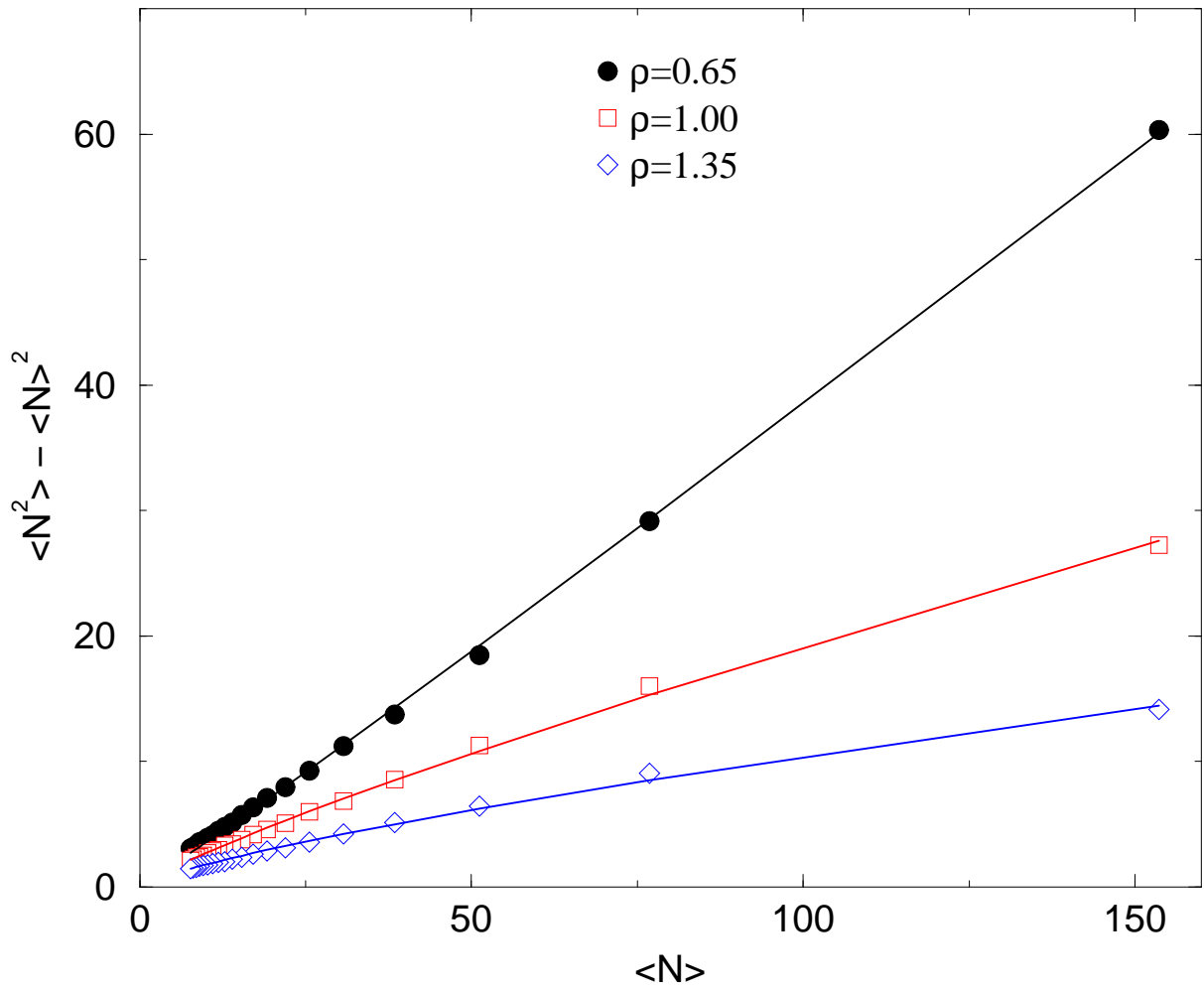
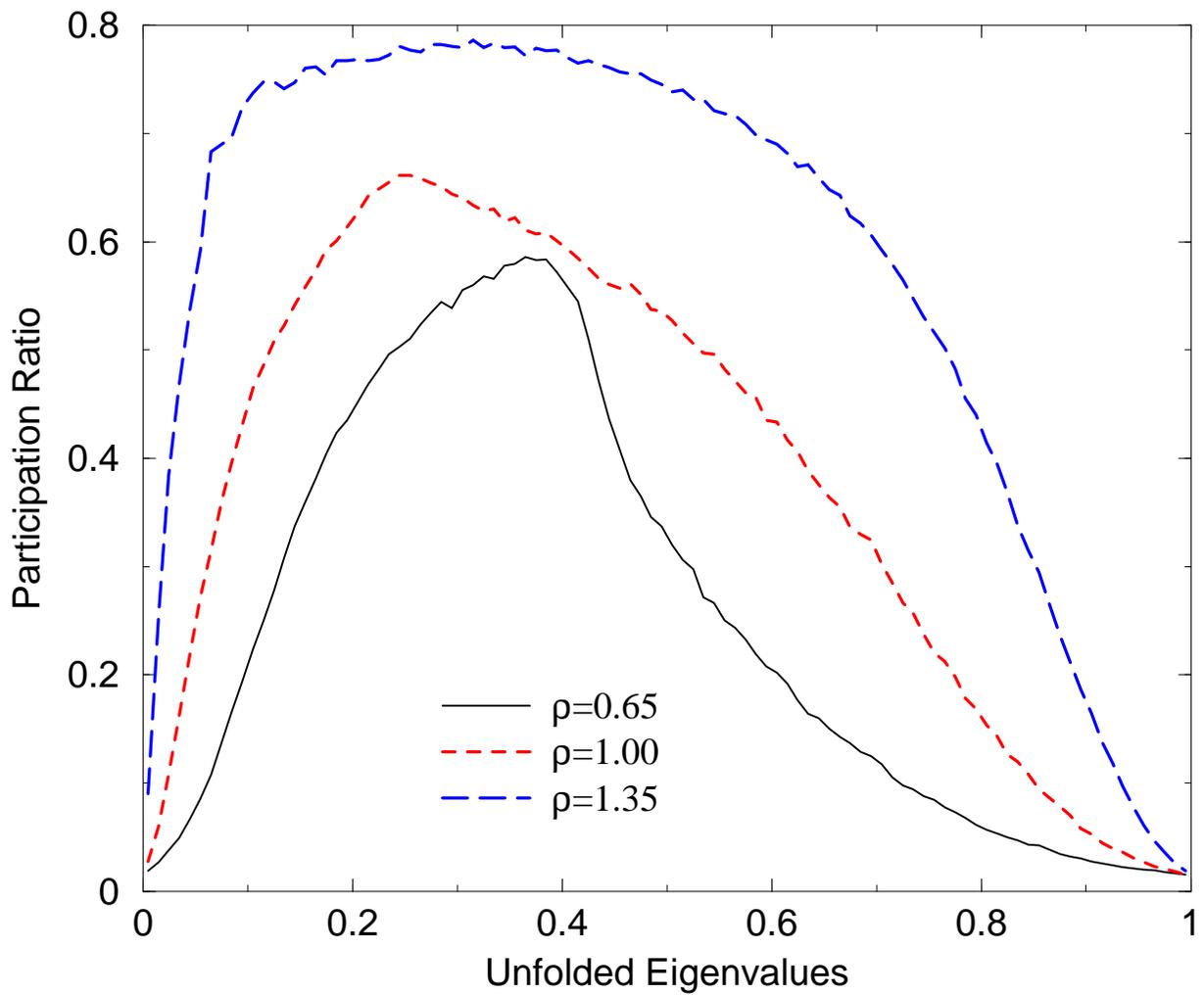
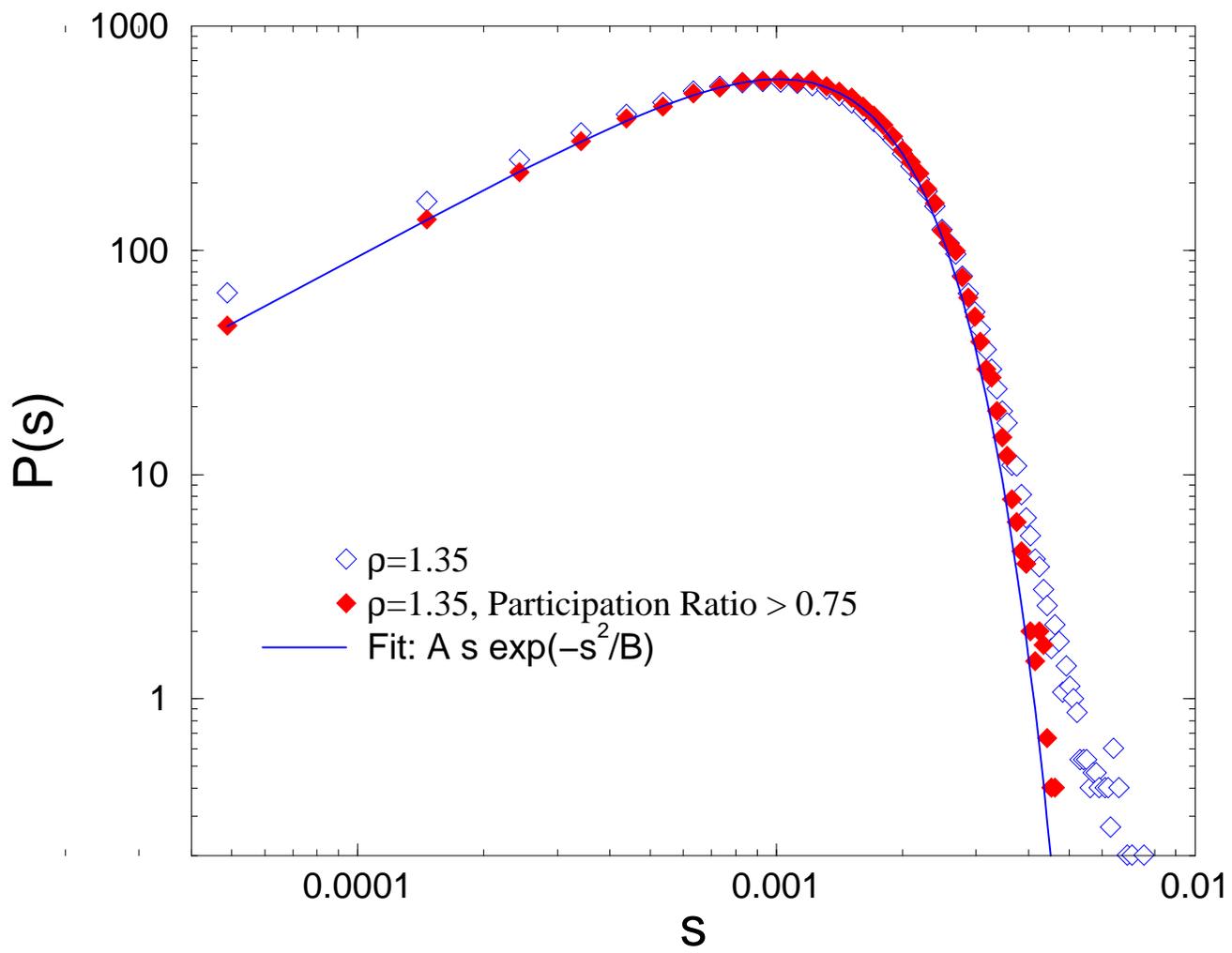
\end{document}